\documentclass[a4paper]{jpconf}
\usepackage{graphicx}
\usepackage{iopams}
\pdfoutput=1
\def\schpt{S\raise0.4ex\hbox{$\chi$}PT}
\def\chpt{\raise0.4ex\hbox{$\chi$}PT}

\begin{document}
\title{Lattice field theory applications in high energy physics}

\author{
Steven Gottlieb
}

\address{Department of Physics, Indiana University, Bloomington, IN 47405, USA}

\ead{sg@indiana.edu}

\begin{abstract}
Lattice gauge theory was formulated by Kenneth Wilson in 1974.  In the
ensuing decades, improvements in actions, algorithms, and computers have
enabled tremendous progress in QCD, to the point where lattice calculations
can yield sub-percent level precision for some quantities.  Beyond QCD,
lattice methods are being used to explore possible  beyond the standard model
(BSM) theories of dynamical symmetry breaking and supersymmetry.  
We survey progress in extracting information about the 
parameters of the standard model by confronting lattice calculations 
with experimental results and searching for evidence of BSM effects.
\end{abstract}

\section{Introduction}
In 1974, Kenneth Wilson invented lattice Quantum Chromodynamics (QCD), a 
non-perturbative approach to Nature's strong force \cite{Wilson:1974sk}.  
Wilson's formulation was
based on using elements of the Lie group $SU(3)$, rather than elements of the 
Lie algebra, which is used in the continuum formulation of the theory.  This
approach allowed Wilson to exactly preserve gauge invariance, which was not
possible when formulating the theory in terms of finite difference operators
applied to elements of the Lie algebra.  Gauge invariance is familiar to us from
electromagnetism, but in QCD it is much richer, as it is based on the
non-Abelian group $SU(3)$, not the Abelian group $U(1)$.  Gauge symmetry 
determines three of Nature's forces: electromagnetic, strong, and weak.

In the years since Wilson's initial paper, which discussed quark confinement
based on a strong coupling expansion, there have been monumental advances in
algorithms, formulations of the theory, and computer power.  In the early
days, it was necessary to neglect the contribution of quantum fluctuations
related to quark-antiquark production and annihilation in the vacuum.
This is called the quenched approximation 
and cannot be systematically improved.
However, it is now possible to include the quantum fluctuations of the
four lightest quarks: up, down, strange, and charm.  
The still heavier bottom and top quarks have negligible effect 
at current precision.
We can even make the up and
down quarks as light as in Nature, which has traditionally been a
difficult computing challenge.  Work on lattice QCD has progressed to the
stage that a number of interesting quantities can be calculated to sub-percent
level, and there have been predictions of particle masses and decay 
properties, not just postdictions. 

Lattice QCD is now extensively used in theoretical studies of elementary
particle and nuclear physics.  The proton-neutron mass difference, the
spectrum of excited baryons, parton moment 
distributions, and properties of light nuclei have all been studied
with varying degrees of precision.  The
properties of QCD at non-zero temperature have been studied to understand
the quark-gluon plasma produced in heavy-ion collisions.
In particle theory, the quark masses for up, down, strange, charm, and bottom,
i.e., all except top, have been calculated.  So, has the strong coupling
$\alpha_s$ and many weak decays that are needed to determine the
Cabibbo-Kobayashi-Maskawa (CKM) quark mixing matrix.
Lattice field theory has also been used to study theories with dynamical 
symmetry breaking as an alternative to the spontaneous symmetry
breaking of the simple Higgs boson theory.  As exciting as it was to
discover the Higgs boson at the LHC, knowing the mass of the Higgs
boson does not solve the mysteries of the Standard Model, and we would
dearly love to find evidence of Beyond the Standard Model physics.  This
might come from seeing new particles at the LHC, but it could just as easily
come from observing anomalies in high precision experiments, where
the anomalies come from small interactions caused by new (virtual)
particles that are too heavy to be produced at the LHC.  Lattice QCD has
an important role to play here in determining the elements of the CKM matrix.
Another area of significant recent progress is in the formulation of
lattice theories with supersymmetry.

Unfortunately, I cannot cover all of these topics, so I shall point
the interested reader to the annual International Symposium on Lattice
Field Theory.  The most recent one was held in July, 2015 in Kobe, Japan 
\cite{Proceedings:2015coa}.
There were over a dozen plenary talks relevant to nuclear and particle
physics.  In this talk, I have relied heavily either on results from my
own collaborations, MILC and Fermilab Lattice-MILC, (referred to as FNAL/MILC
below), 
or state-of-the-art summaries prepared by the
Flavor Lattice Averaging Group (FLAG) \cite{Aoki:2013ldr}
an international group of scientists who
critically review work on a large number of quantities in lattice QCD
and prepare averages for ease of use by a wider (non-lattice) community.
The last FLAG review appeared in 2013 
\cite{Aoki:2013ldr} and a new one will appear in early 2016.
I am pleased to be a member of FLAG.  

\section{The Standard Model and lattice QCD}
The Standard Model is a theory of quarks, leptons, gauge bosons, and the Higgs
boson.  It describes only three of the known forces, as gravity is not included.
The model is described by its symmetries and the matter content.  The 
symmetry is $SU(3) \times SU(2) \times U(1)$.  The group $SU(3)$ is the
symmetry of QCD and $SU(2)\times U(1)$ is that of the electroweak
interactions.  Spin-1 particles are the force carriers.  They are called
gluons (for QCD), and the photon and weak bosons (specifically, $W^\pm$ and
$Z$) for the electromagnetism and the weak force, respectively.  
The Higgs boson has no intrinsic spin.  The quarks and
leptons are spin-1/2 matter particles.  The quarks interact with all the
force carriers.  The charged and neutral leptons don't interact with
the gluons, but they do interact with the weak force carriers.  The charged
leptons interact with the photon, but the neutral ones (neutrinos) do not.
One of the reasons we think there is physics beyond the the Standard Model is
that the model has many undetermined parameters.  These parameters must be
determined from experiment (with various inputs from theory).  In a more
fundamental theory, there might be relations between the parameters, so
they would not seem as arbitrary as they do now.

For each of the three symmetry groups there is a coupling constant.  
For $SU(3)$, it is called $g_s$.  For $ SU(2) \times U(1)$, the 
two couplings are $g$ and $g'$.  There are six quark masses.  There
are three masses for the charged leptons.  
Now that we know neutrinos have mass, there are also three neutrino masses.
The Cabibbo-Kobayashi-Maskawa quark mixing matrix (detailed in the next section)
is complex and unitary.  An arbitrary $3\times3$
complex matrix would have 18 real parameters; however, because of unitarity
and our ability to chose some phases of the quark fields, there are only four
independent parameters that determine the CKM matrix.  These are commonly 
described as three angles and a complex phase factor that determines
CP violation.  
The combination of discrete symmetries charge conjugation C, and parity 
is denoted by CP.
There is a similar matrix for the neutrinos called PMNS for Pontecorvo, Maki,
Nakagaw and Sakata.  However, since the neutrinos do not interact strongly,
we will have no more to say about that.  Lattice QCD input is important
for determination of ten parameters of the Standard Model:  $\alpha_s=g_s^2/(4\pi)$,
the four parameters that determine the CKM matrix, and $m_u$, $m_d$, $m_s$,
$m_c$, and $m_b$.  The sixth quark, the top quark, decays weakly before it can
form bound states, so we do not need lattice QCD to study its mass.

Lattice QCD provides a nonperturbative treatment of the quantum field theory
that describes the strong interaction.  Because the coupling is strong,
many phenomena cannot be calculated perturbatively.
Quantum field theories require regularization and renormalization. 
The lattice technique provides one such regularization.
However, numerical errors must be carefully controlled. 
Errors come from the non-zero lattice spacing (continuum limit),
finite volume (infinite volume limit), and
unphysical light quark masses (chiral extrapolation).  In parentheses, we
have the limit or operation that must be done to control the systematic
error from each effect.  In addition, there are statistical errors.
Groups are increasingly able to work with up and down quark
masses very close to their physical value, which greatly reduces errors from
the chiral extrapolation that were seen in earlier calculations.
There are at least five popular ways to deal with the quarks in lattice QCD.
They go by the names: Wilson/Clover, staggered, domain wall, twisted mass,
and overlap.  Each method has different systematic errors at nonzero lattice 
spacing, so it is useful to use different methods and compare the final
results after all errors are controlled.
The number of dynamical flavors used also varies by collaboration.
The most phenomenologically relevant calculations use dynamical up, down, 
and strange quarks, or those plus charm.  
These are denoted $N_f=2+1$ or $2+1+1$, 
respectively, because the up and down quarks are usually treated as if
their masses were identical.  (Their average mass is used.)

\section{CKM matrix}
It has been observed for many years that the Universe contains much more matter 
than antimatter.  This is known as the baryon asymmetry.
Kobayashi and Maskawa won the Nobel prize for their realization that with 
three (or more) generations we can have CP violation, which might 
explain the baryon asymmetry of the Universe.  However, we now know that the CP
violation in the strong interaction is probably too weak for this
purpose, and it may be the CP violation appearing in the PMNS matrix for
neutrino mixing that accounts for the baryon asymmetry.  Here is the CKM mixing
matrix (bold notation) augmented with some of the decay or mixing processes
that can be used to determine each matrix element:
\begin{equation}
\left( 
\begin{array}{c c c}
\mathbf{V_{ud}} & \mathbf{V_{us}} & \mathbf{V_{ub}} \\
\pi \rightarrow l\nu & K\rightarrow l\nu  & B\rightarrow l\nu\\
                     & K\rightarrow \pi l\nu & B\rightarrow \pi l\nu \\
\mathbf{V_{cd}} & \mathbf{V_{cs}} & \mathbf{V_{cb}} \\
D\rightarrow l\nu & D_s\rightarrow l\nu & B_c\rightarrow l\nu \\
D\rightarrow \pi l\nu & D\rightarrow K l\nu & B \rightarrow D^{(*)} l\nu \\
\mathbf{V_{td}} & \mathbf{V_{ts}} & \mathbf{V_{tb}} \\
B_d \leftrightarrow \bar B_d & B_s \leftrightarrow \bar B_s & \end{array}
\right) .
\end{equation}
In the second and fifth rows, we have meson decays called leptonic, because only
a charged lepton and a neutrino appear in the final state.  The third and six
rows contain decays called semi-leptonic because there is also a meson in
the final state.  The last row contains two meson mixing processes that
determine the CKM matrix elements just above them.
Since the CKM matrix is unitary, each row and each column is a complex unit
vector.  Also, each row (column) is orthogonal to the other rows (columns)
leading to the so-called unitarity triangle in the complex plane.
Violations of unitarity are evidence of non-standard-model physics.
Further, if two different processes are used to determine an element of 
the matrix and they do not agree, that is evidence for new BSM physics,
which we would dearly love to find.  We will examine both these tests
of the SM.

If we could do experiments on free quarks, it would be easy to 
determine mixing; however, confinement means we need to deal with bound states.
Thus, LQCD input for decay constants and form factors is needed to determine 
elements of the CKM matrix.  For example, the branching fraction
for the leptonic decay of a $D_{(s)}$ meson is given by
\begin{equation}
{\mathcal{B}}(D_{(s)} \to \ell\nu_\ell)= {{G_F^2|V_{cq}|^2 \tau_{D_{(s)}}}\over{8 \pi}} f_{D_{(s)}}^2 m_\ell^2 
m_{D_{(s)}} \left(1-{{m_\ell^2}\over{m_{D_{(s)}}^2}}\right)^2
\end{equation}
where the unknowns are $|V_{cq}|$, the absolute value of the CKM matrix 
element with $q=d$ or $q=s$ for the $D$ or $D_s$ meson, respectively, and
$f_{D_{(s)}}$ is the corresponding decay constant, which needs to be calculated
in LQCD.  The other quantities, such
as masses, lifetimes and the Fermi constant are easily found from experiment.

\subsection{Light quarks and the first row}
\begin{figure}[th]
\begin{minipage}{75mm}
\includegraphics[width=75mm]{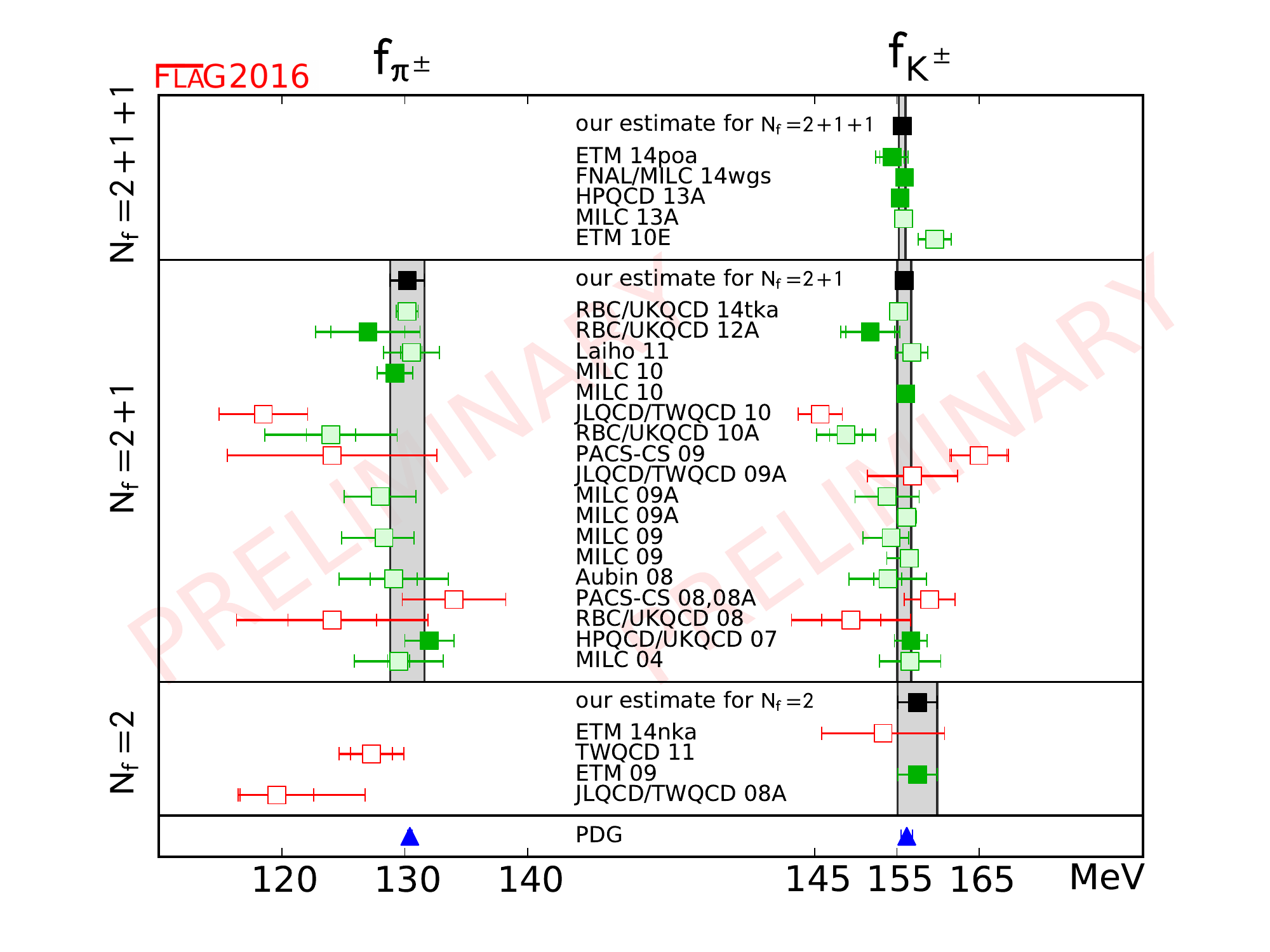}
\end{minipage}\hspace{9mm}%
\begin{minipage}{75mm}
\includegraphics[width=75mm]{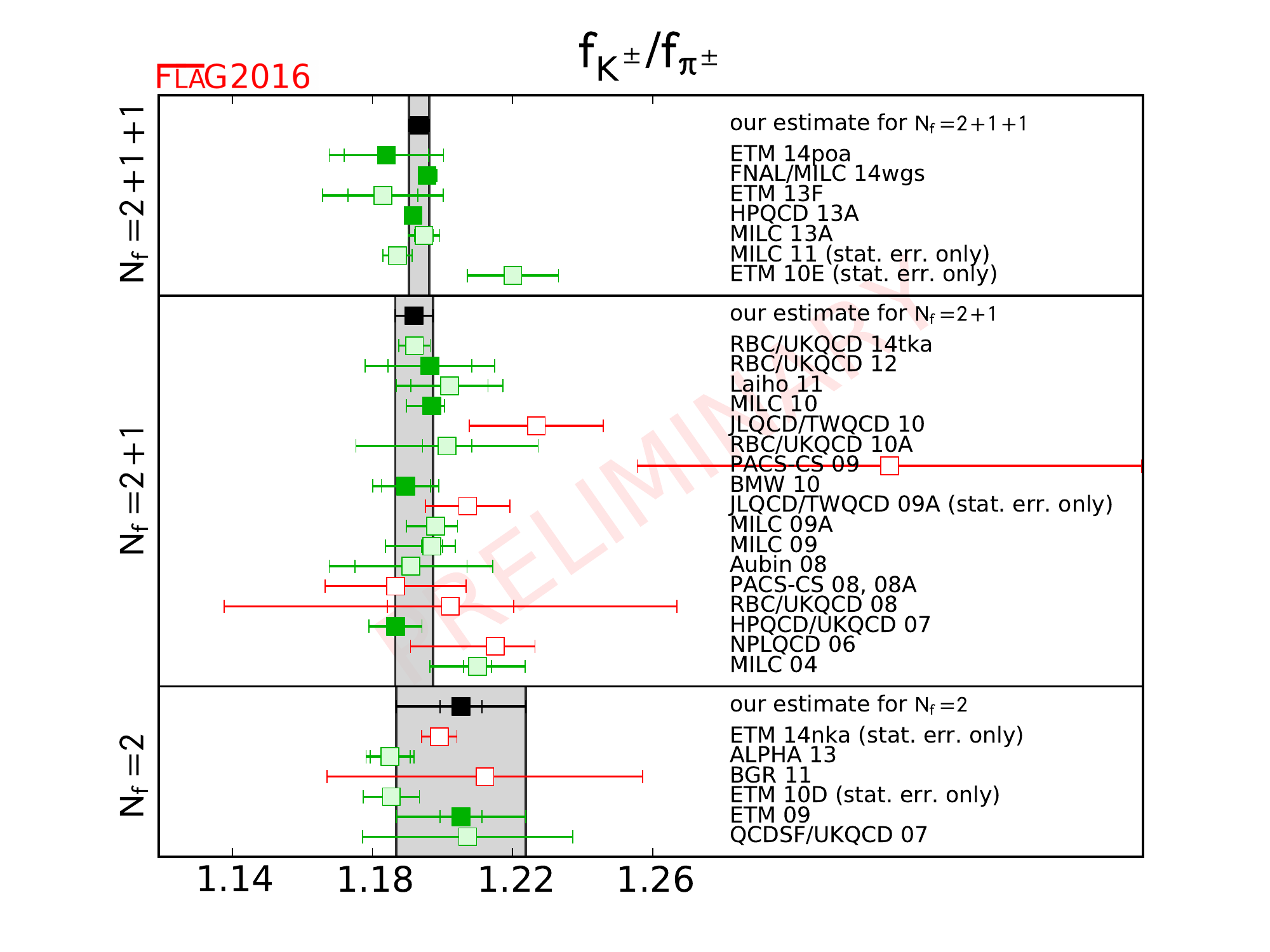}
\end{minipage}
\caption{\label{fig:lightdecay}
(L) FLAG \cite{Aoki:2013ldr} summary of LQCD results for the charged 
pion and kaon decay
constants.  (R) The ratio of charged kaon to charged pion decay constants.
The meaning of the colors, symbols and bands is explained in the text.
}
\end{figure}

We will begin our discussion with results for mesons that contain only the
three lightest quarks up, down, and strange.  The ground states are called
pions and kaons.
In Fig.~\ref{fig:lightdecay}(L), we see the FLAG summary of LQCD results stretching back over a decade by various
groups.  Results in red are deemed to have insufficient
control of all systematic errors.  Results with a solid green symbol are 
included in the FLAG estimate (black point with error bar and vertical gray
bands).  Points with light green have been superseded or not been refereed,
hence not included in the FLAG estimate.  Blue points come  from the Particle Data Group (PDG) \cite{Agashe:2014kda}
or a non-LQCD method.  Calculations with different numbers of dynamical quarks
are considered separately.  In some cases, $f_\pi$ has been used to set the
lattice spacing (or scale), so only $f_K$ is shown.  We see excellent agreement
with the values from the PDG.
The ratio $f_K/f_\pi$
can be calculated accurately and used to determine $|V_{us}/V_{ud}|$ from
precise measurements of the ratio of pion and kaon decay rates which show that
$\left|{V_{us}\over V_{ud}}\right| {f_{K^\pm}\over f_{\pi^\pm}} = 0.2758(5)$.
Figure~\ref{fig:lightdecay}(R) shows results for the decay constant ratio.

\begin{figure}[t]
\begin{minipage}{50mm}
\includegraphics[width=50mm]{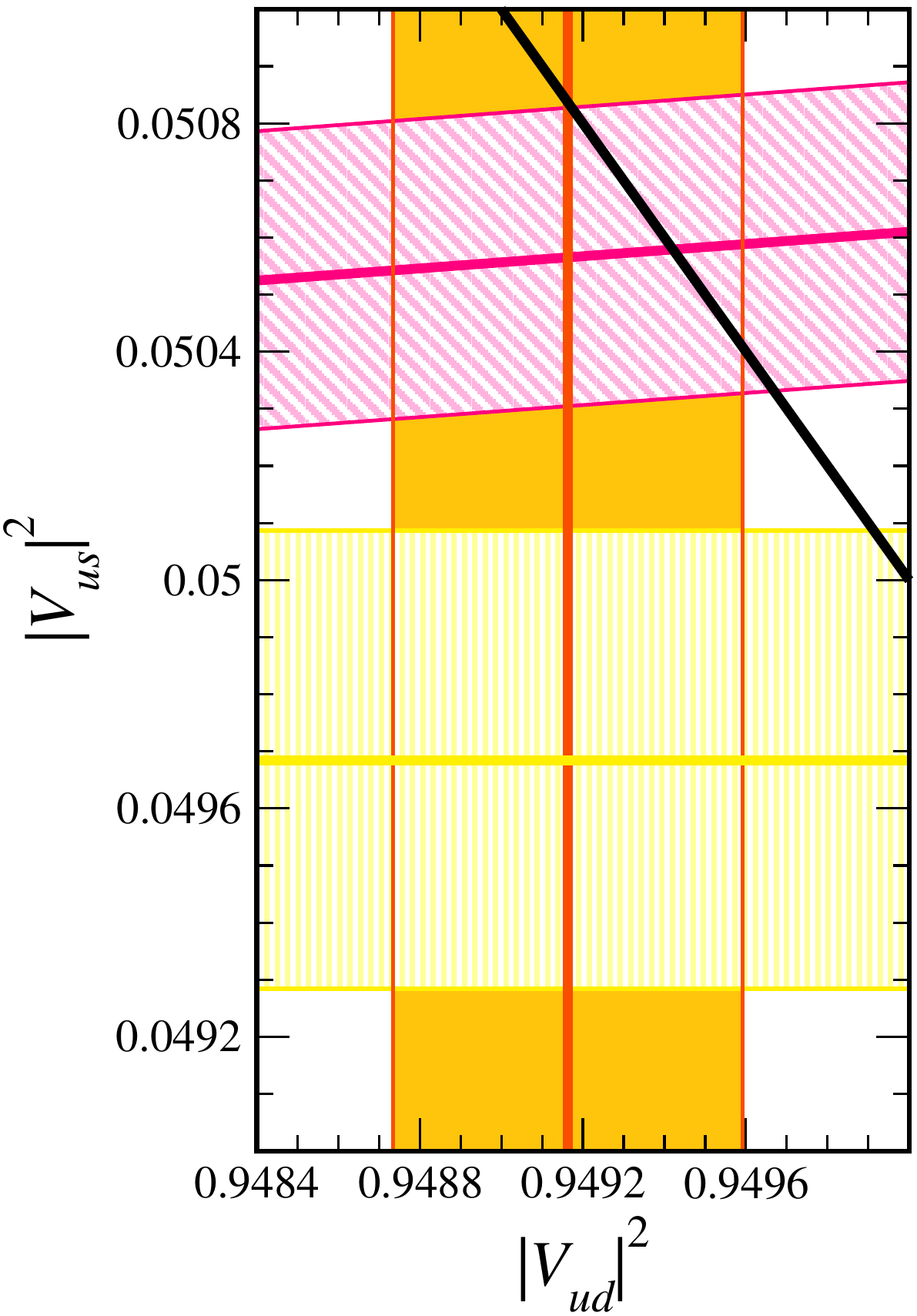}
\end{minipage}\hspace{9mm}%
\begin{minipage}{100mm}
\includegraphics[width=100mm]{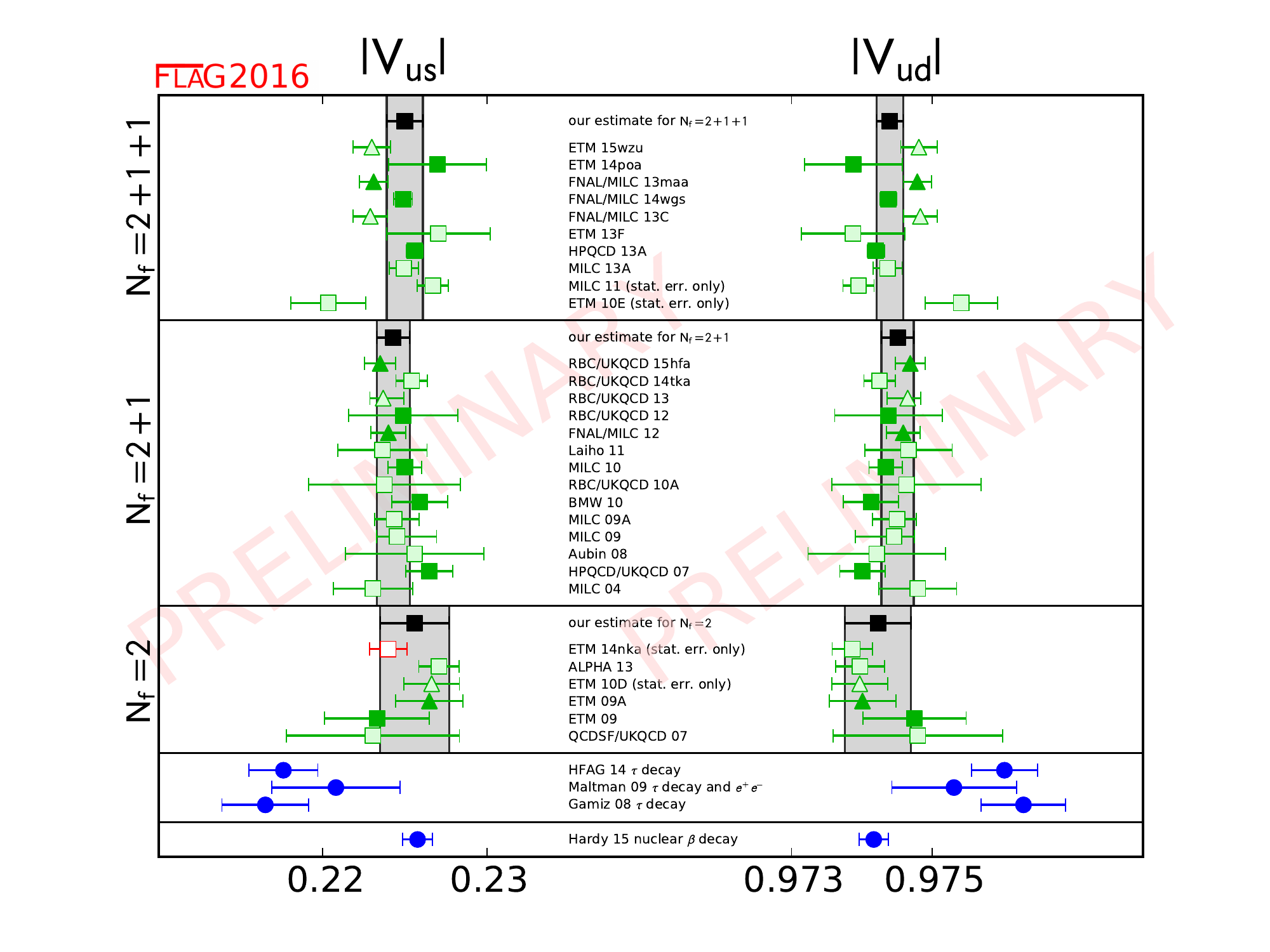}
\end{minipage}
\caption{\label{fig:firstrow}
(L) Unitarity test for the first row of the CKM matrix from 
Ref.~\cite{Bazavov:2014wgs}.  Black line shows unitarity constraint.  Colored
bands explained in text.  (R) Subset of FLAG \cite{Aoki:2013ldr}
summary for $|V_{us}|$ and $|V_{ud}|$.  Most red points eliminated for readability.
}
\end{figure}

Let's turn to the semileptonic kaon decay.  
Semileptonic decays have three-body final states, so there is one kinematic variable, usually denoted $q^2$, which is the square of the momentum transfer to the leptons.
From 4-momentum conservation, $p_K = p_\pi + q_l +q_\nu$ and $q = q_l + q_\nu$,
where we have used $p$ for hadron momenta and $q$ for lepton momenta, with
the subscript denoting the particle.
To extract $|V_{us}|$, we just need the form factor at zero momentum transfer,
i.e., $f_+(0)$ as experiment tells us that 
$|V_{us}| f_+(0) = 0.2163(5)$.  This can be combined with the FNAL/MILC
$N_f=2+1+1$ result \cite{Bazavov:2013maa}
$ f_+(0) = 0.9704(24)(220$ to determine an
error band for $|V_{us}|$.
First row unitary states $|V_{ud}|^2 + |V_{us}|^2 + |V_{ub}|^2= 1$.
However, as we will see below $|V_{ub}|\approx 4\times 10^{-3}$, so the
last term can be neglected as the errors on the first two terms are
a few times $10^{-4}$.  Thus, the unitarity constraint will be a straight
line in the $|V_{ud}|^2$ - $|V_{us}|^2$ plane.
In Fig.~\ref{fig:firstrow}(L), we show the unitary constraint as a black
line, along with the angled error band from leptonic pion and kaon decay,
the horizontal error band from kaon semileptonic decay, and a vertical
error band from nuclear $\beta$-decay that is independent of LQCD calculations.
We see that there is some tension between the two types of decay studied
in LQCD, but that unitarity, leptonic decays, and  $\beta$-decay are in good
agreement.

A summary of many determinations of $|V_{ud}|$ and $|V_{us}|$, based on either
leptonic or semileptonic decays, has been prepared by FLAG.  This is shown
in Fig.~\ref{fig:firstrow}(R) where squares indicate leptonic decays 
and triangles indicate semileptonic.  The blue points, as usual, are from
non-lattice calculations.  Careful inspection of the $N_f=2+1+1$ 2014
results from MILC and FNAL/MILC shows the tension we have seen above
between leptonic and semileptonic decays.  Other calculations do not yet
have the precision to confirm the discrepancy or rule it out.  The errors
on the FLAG estimates for $N_f=2$ and $2+1$ are larger than for $2+1+1$ 
and they do not have a tension with unitarity.  The FLAG estimate for
$2+1+1$ does show some tension with unitarity.  First row unitarity
is not the first place we would expect to find evidence for BSM flavor
physics, so it will be interesting to improve these calculations, particularly
to calculate the kaon semileptonic form factor over its entire kinematic
range.  There is also the interesting tension between the results near
the bottom of the figure, based on $\tau$ decays, and those for pion and kaon
decay.

\subsection{Charm decays and the second row}
\begin{figure}[th]
\begin{minipage}{100mm}
\includegraphics[width=100mm]{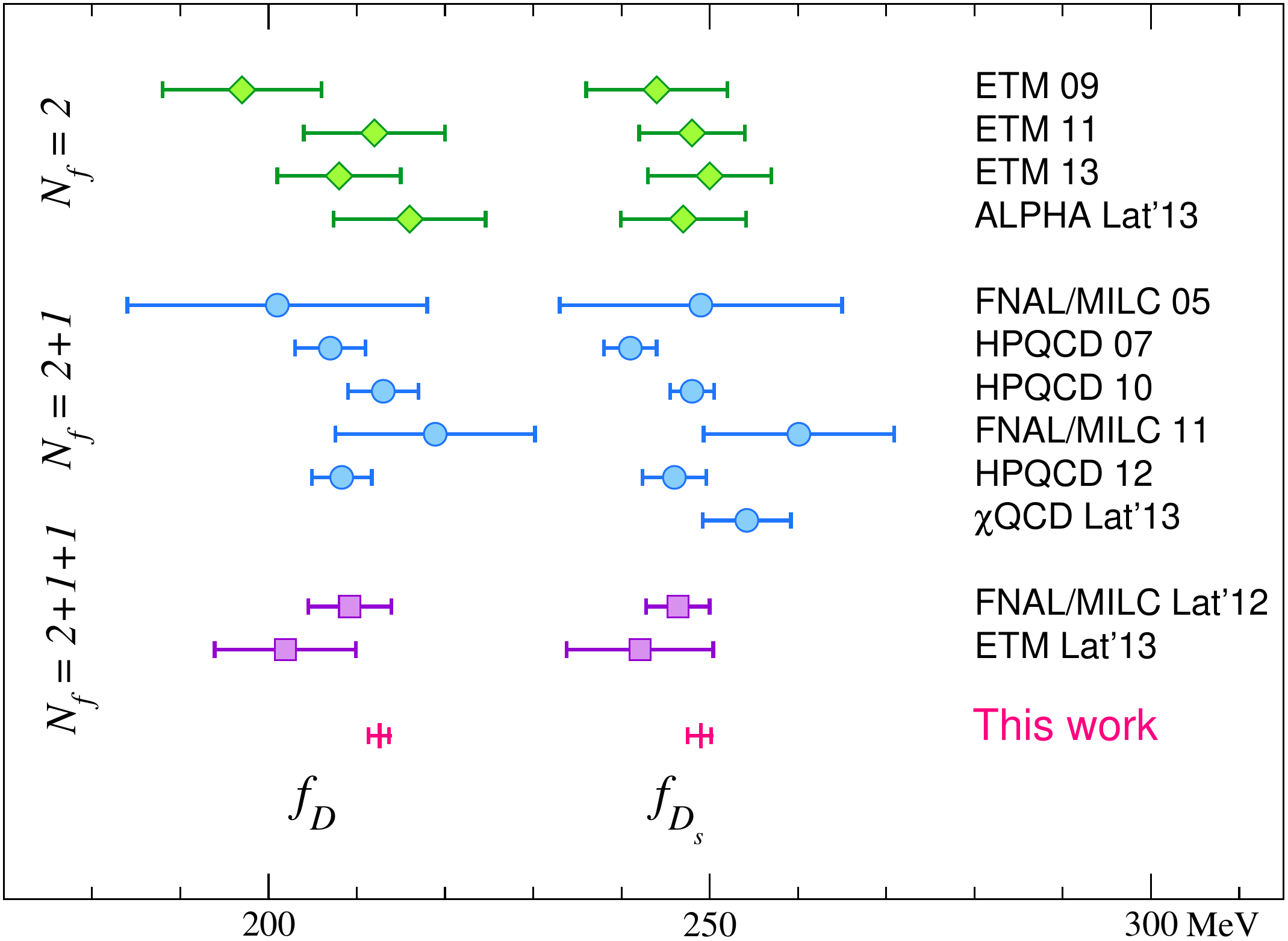}
\end{minipage}\hspace{9mm}%
\begin{minipage}{50mm}
\includegraphics[width=50mm]{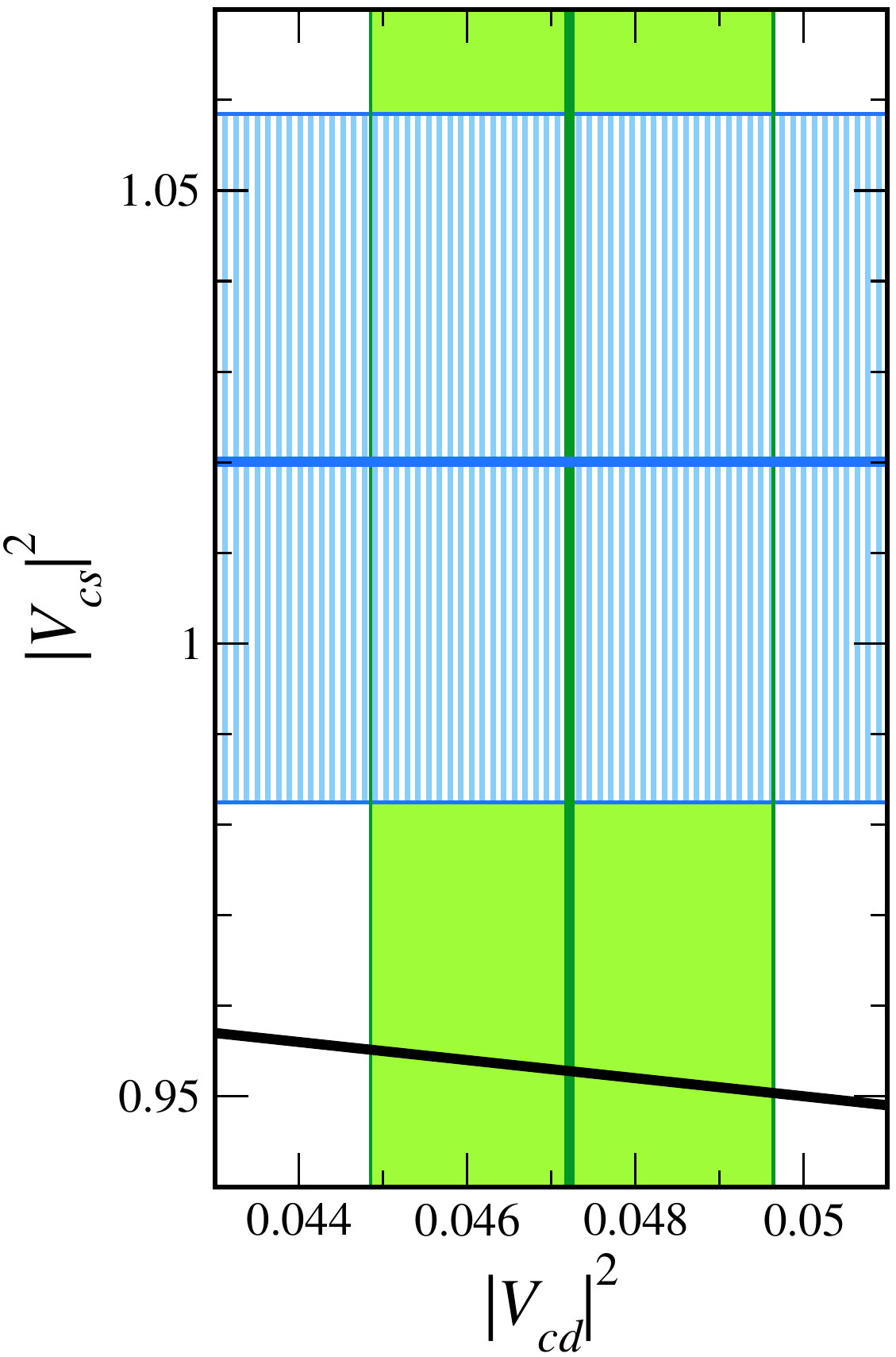}
\end{minipage}
\caption{\label{fig:secondrow}
(L) Summary of results for the charm meson decay constant prepared by 
FNAL/MILC \cite{Bazavov:2014wgs}.  ``This work'' refers to the calculation
presented there.
(R) Unitarity test for the second row of the CKM matrix from 
Ref.~\cite{Bazavov:2014wgs}.  
}
\end{figure}

Leptonic and semileptonic decays of the $D$ and $D_s$ mesons have been
studied on the lattice, but not as extensively as for the pion and kaon.  It has
been about a decade since decay constant predictions of FNAL/MILC were
tested at CLEO-c \cite{Artuso:2005ym}.  Initial errors were about 10\%, but
current errors from FNAL/MILC are only 0.6\%.  A great deal of the improvement
is due to the use of highly improved staggered quarks (HISQ) that were developed
by the HPQCD collaboration.  Figure~\ref{fig:secondrow}(L) compares the 2014
FNAL/MILC results (labeled ``This work'') with prior calculations, mostly
by the European Twisted Mass collaboration and HPQCD.  The FNAL/MILC
results are $f_{D^+} = 212.6(0.4)({}^{+1.0}_{-1.2})\ \mathrm{MeV}$,
$f_{D_s} = 249.0(0.3)({}^{+1.1}_{-1.5})\ \mathrm{MeV}$, and 
$f_{D_s}/f_{D^+} = 1.1712(10)({}^{+29}_{-32})$ for the decay constant ratio,
for which there is some cancellation of the systematic errors.
For references to the other work and the HISQ action see
Ref.~\cite{Bazavov:2014wgs}.

To make use of these decay constants, we rely on the work of 
Rosner and Stone \cite{Rosner:2013ica}
to summarize the experimental results.
They find $f_{D}|V_{cd}|=46.06(1.11)  \mathrm{MeV}$, and 
$f_{D_s}|V_{cs}|=250.66(4.48) \mathrm{MeV}$.  
The experimental errors are 1.8--2.4\%.

Combining the experimental results and the FNAL/MILC decay constants
gives
$|V_{cd}|=0.217(1) (5)(1)$, and $|V_{cs}|= 1.010(5)(18)(6)$, where
the errors are lattice, experiment and structure-dependent
electromagnetic, respectively.  Thus, the experimental errors are currently 
dominant.  In Fig.~\ref{fig:secondrow}{R}, we see evidence for an $\approx 1.8
\sigma$ tension with unitarity for the two leptonic charm decays.  The black
line is the unitarity constraint.  The horizontal blue band is for $D_s$
decay and the vertical green band is for $D^+$ decay.  Once again, the
third element of the row $V_{cb}$ is too small to make a difference
at the current level of precision.  The semileptonic form factors for
$D_{(s)}$ mesons are much less studied than for light quarks; however,
there should be some updates in the coming year.  We refer the reader
to FLAG  \cite{Aoki:2013ldr} for details.

\subsection{$B$ meson decays}
The $b$ quark is the heaviest one that forms bound states that can be studied
with LQCD.  Both leptonic and semileptonic decays of $B$ and $B_s$
mesons have been studied.  In addition to the usual decays that produce
a charged lepton and a neutrino, there are a number of rare decays
that involve the so-called flavor changing neutral current (FCNC).
In the SM, the FCNC vanishes at the tree level, so small quantum loop effects
from new physics may be visible.  That makes the rare decays a promising
topic to study.  In fact, there have been some recent results from the
LHCb experiment at the Large Hadron Collider that show tensions with
the SM predictions.  These rare decays also present an alternative way to
determine $|V_{td}|$ and  $|V_{ts}|$ that can be compared with the
meson mixing processes indicated in our CKM matrix.

FLAG has summarized results for decay constants $f_B$ and 
$f_{B_s}$ \cite{Aoki:2013ldr}.
The errors on these decay constants are about 2\% for $N_f=2+1$ and
$2+1+1$.  For  $N_f=2+1$, $f_B=190.5(4.2) \mathrm{MeV}$ and
$f_{B_s}=227.7(4.5) \mathrm{MeV}$.  
Unfortunately only $B\rightarrow \tau\nu$
has been observed so far and the error is about 20\%.  
So, in this case the LQCD calculation is ahead of the measurement.
This allows a determination of $|V_{ub}|$, but it is not competitive with
the value from semileptonic decay.

\begin{figure}[th]
\begin{minipage}{75mm}
\includegraphics[width=75mm]{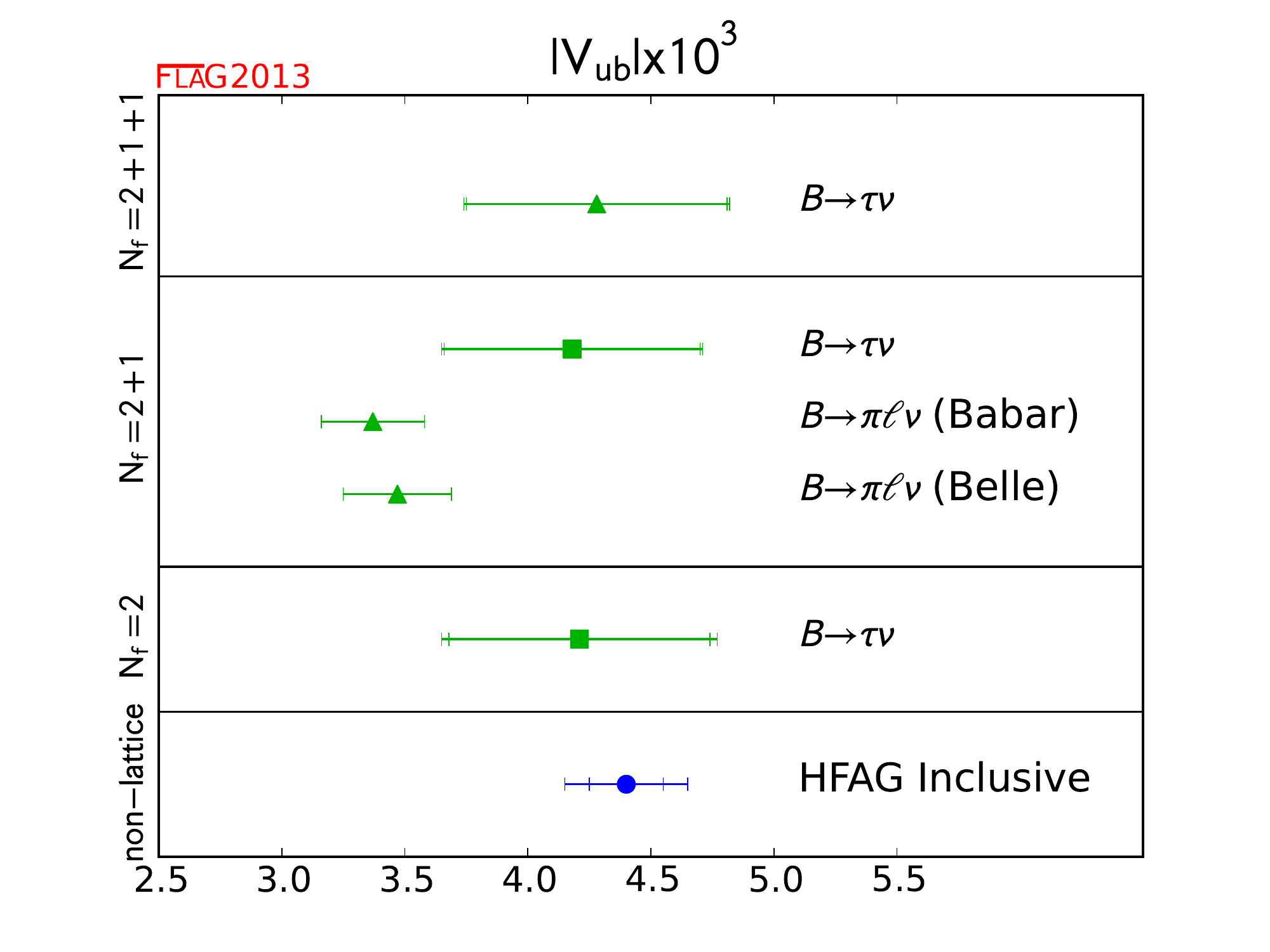}
\end{minipage}\hspace{4mm}%
\begin{minipage}{85mm}
\includegraphics[width=85mm]{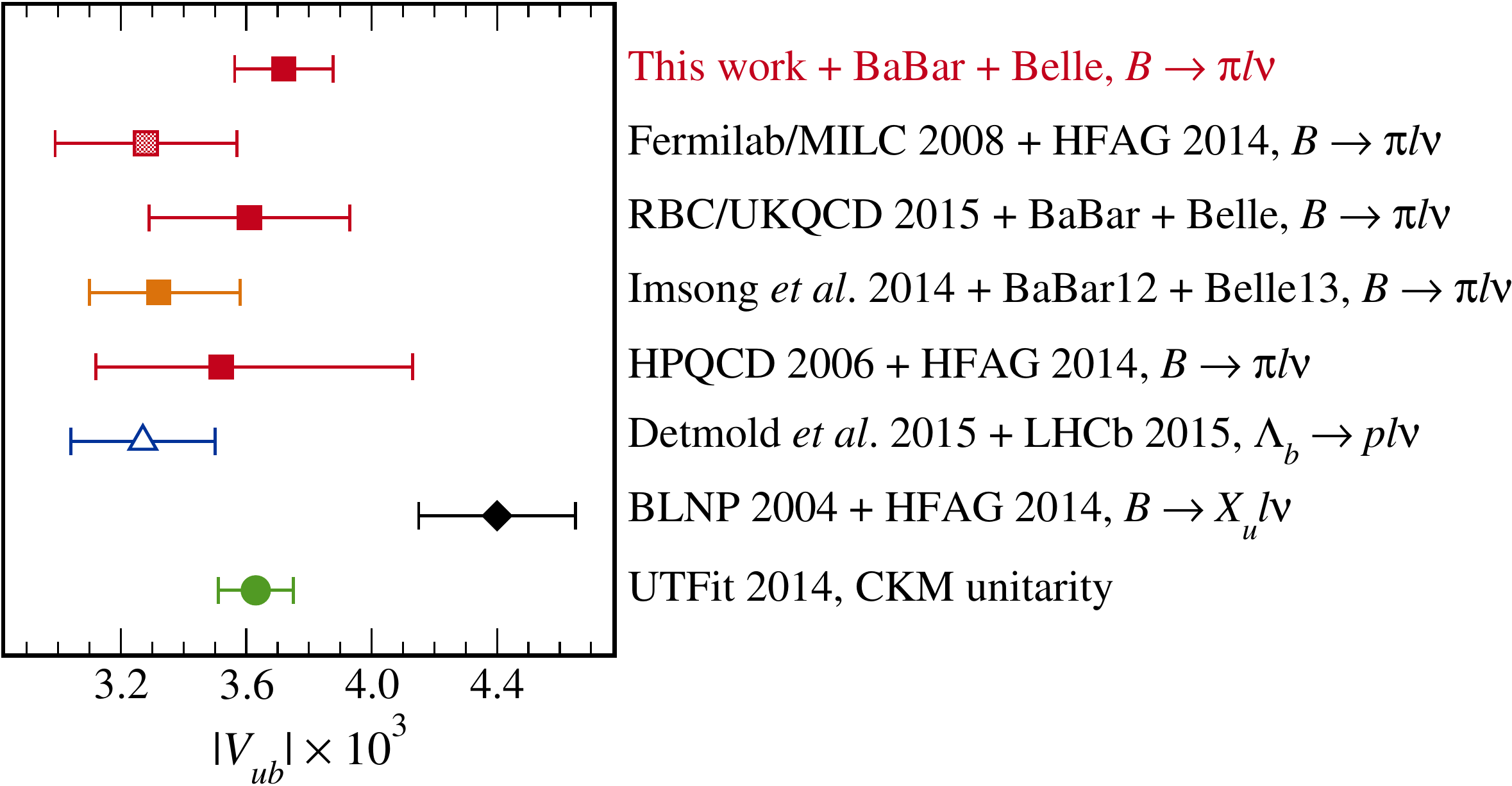}
\end{minipage}
\caption{\label{fig:bdecayconstants}
(L) FLAG \cite{Aoki:2013ldr} summary for $|V_{ub}|$ including results for
$N_f=2$, 2+1, and 2+1+1, as well as the inclusive result.
(R) FNAL/MILC \cite{Lattice:2015tia} summary of results for 
$|V_{ub}|$, including
determination from $\Lambda_b$ decay.  See Ref.~\cite{Lattice:2015tia} for
original sources.
}
\end{figure}

The semileptonic decays $B\rightarrow \pi \ell\nu $ and  
$B_s\rightarrow K \ell\nu $ have been studied on the lattice.  The former has
been observed at BaBar and Belle, but the latter has not been observed.
Another way to determine $|V_{ub}|$ is from inclusive decays.  There is
a long standing tension between that determination and the one from 
$B\rightarrow \pi \ell\nu $.  The central value of $|V_{ub}|$
based on the SM analysis of the leptonic decay is between that from
the semileptonic exclusive decay and the inclusive method.  However,
its error bar, limited by experiment, is too large to help clarify
the situation.  Belle II will improve the $B\rightarrow \tau\nu$ measurement,
which should really help resolve these issues.
Figure~\ref{fig:bdecayconstants}(L) shows the FLAG summary from 2013.
There are new results for $B\rightarrow \pi \ell\nu $ from
FNAL/MILC and RBC/UKQCD.  There is also a new determination
of $|V_{ub}|$ based on decay of the $\Lambda_B$ baryon recently seen
at LHCb.  These determinations are compared in 
Fig.~~\ref{fig:bdecayconstants}(R) \cite{Lattice:2015tia}.
FLAG computed values of $|V_{ub}|$ based on the BaBar and Belle experiments.
They found $|V_{ub}|=3.37(21) \times 10^{-3}$ and $3.47(22)\times 10^{-3}$,
respectively, based on 2+1 flavor LQCD.  The new FNAL/MILC result
which uses results from both BaBar and Belle is $3.72(16) \times 10^{-3}$
which reduces, but does not eliminate the tension between exclusive and
inclusive decays.

\begin{figure}[th]
\begin{minipage}{75mm}
\includegraphics[width=85mm]{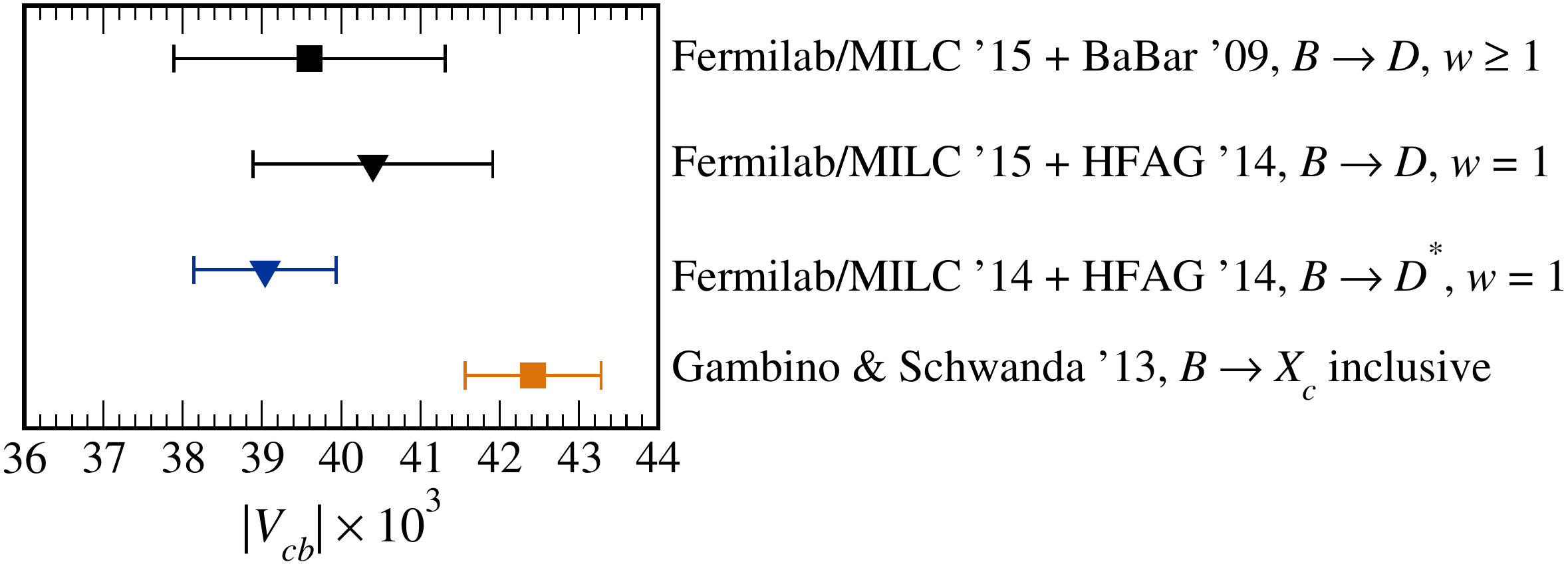}
\end{minipage}\hspace{14mm}%
\begin{minipage}{65mm}
\includegraphics[width=65mm]{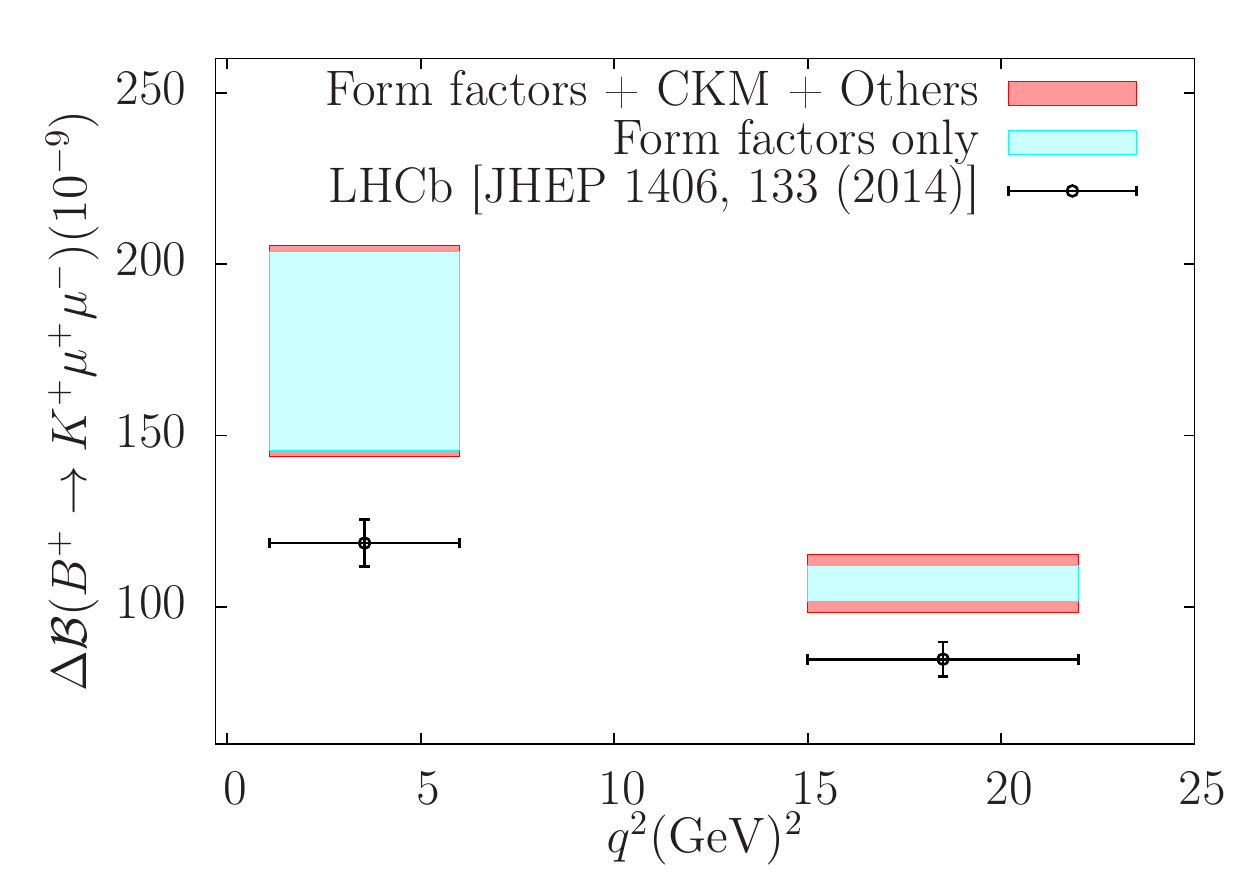}
\end{minipage}
\caption{\label{fig:vcbdecays}
(L) FNAL/MILC \cite{Lattice:2015rga} summary of results for $|V_{cb}|$, 
including determination from inclusive $B$ decays to $D$ or $D^*$, and 
inclusive decays.
(R) Branching fraction for $B^+\rightarrow K^+\mu^+\mu^-$ showing tension 
between SM prediction \cite{Du:2015tda} and recent LHCb measurement.

}
\end{figure}

The last entry in the second row of the CKM matrix, $V_{cb}$, can be studied
in the exclusive decays $B\rightarrow D^*\ell\nu$ and $B\rightarrow D\ell\nu$.
It can also be determined in inclusive decays where the decay products must
include charm quarks.  FNAL/MILC has recently compared $|V_{cb}|$ based
on both determinations and there is again some tension between inclusive
and exclusive results.  As can be seen in Fig.~\ref{fig:vcbdecays}(L), the errors
from the decay to $D^*$ are somewhat smaller than that from decay to
$D$.  With current errors, those decays agree with each other reasonably
well and the real tension is between $B\rightarrow D^*\ell\nu$ and the
inclusive value of $|V_{cb}|$.

Turning to rare $B$ decays, FNAL/MILC has recently calculated the form factors
needed for both SM and BSM decays through a 
FCNC \cite{Bailey:2015nbd,Bailey:2015dka}.  As mentioned above,
this is a promising place to look for new physics.  There is some tension
between the SM prediction and recent LHCb measurements of
$B^+\rightarrow \pi^+\mu^+\mu^-$ and  $B^+\rightarrow K^+\mu^+\mu^-$.
The LHCb measurement is smaller than the SM prediction in three of four
fairly wide bins of $q^2$, the square of the momentum transfer to the muons.  
Figure~\ref{fig:vcbdecays}(R) shows the comparison for 
$B^+\rightarrow K^+\mu^+\mu^-$ where the difference is more pronounced.
In Ref.~\cite{Du:2015tda}, both processes are shown.
For all four bins for the two processes, the tension is $1.7\sigma$.

\section{Conclusions}
Much progress has been made in lattice QCD, and more generally
in lattice field theory.  We concentrate here on quantities
needed for the study of the CKM matrix.  Calculational precision
is now high enough that we can begin to look for evidence of BSM physics.
We have seen a number of tensions
between 1.5 and 2 standard deviations related to the CKM matrix.  It will be
interesting to see if reduced errors from both theory and experiment result
in stronger hints (or perhaps significant evidence) of BSM physics.
In the oral presentation, quark
masses and $\alpha_s$ were also discussed.  (See Ref.~\cite{Aoki:2013ldr} for
details.)
\ack
I thank all my colleagues in the MILC (especially C.B.) and 
Fermilab Lattice Collaborations, as well as all the current and past members
of the Flavor Lattice Averaging Group.  It has been a real pleasure working with them
to produce new results and to make it easy for others to follow the 
progress in LQCD.  This work was supported by the US Department of Energy
grant DE-SC0010120.  I also thank Profs. Santra and Ray for organizing such
a wonderful conference, inviting me to attend, providing great hospitality,
and helping to support my visit.

\section*{References}
\bibliography{gottlieb_ccp2015.bib}
\end{document}